\documentclass[pra,amsmath,amssymb,twocolumn,superscriptaddress]{revtex4}

\usepackage{amsmath,amssymb}
\usepackage[usenames]{color}
\usepackage{amssymb}
\usepackage{grffile}
\usepackage[pdftex]{graphicx}
\usepackage{amsmath, amstext, amssymb, amsfonts, amsxtra}
\usepackage{textcomp}
\usepackage{xspace}
\usepackage{url}
\usepackage[colorlinks,linkcolor=red,anchorcolor=blue,citecolor=blue,urlcolor=blue]{hyperref}

\newcommand{\beq}{\begin{equation}}
\newcommand{\eeq}{\end{equation}}
\newcommand{\beqa}{\begin{eqnarray}}
\newcommand{\eeqa}{\end{eqnarray}}

\definecolor{myorange}{RGB}{255,117,40}

\newcommand{\im}{{\rm i}}
\newcommand{\fid}{\mathcal{F}}
\DeclareMathOperator{\tr}{tr}

\begin{document}
\title{Towards generation of cat states in trapped ions set-ups via FAQUAD protocols and dynamical decoupling}
\author{Mikel Palmero}
\email{palmerolazcoz@sutd.edu.sg}
\affiliation{Science and Math Cluster, Singapore University of Technology and Design, 8 Somapah Road, 487372 Singapore}
\author{Miguel \'Angel Sim\'on}
\affiliation{Departamento de Qu\'imica-F\'isica, Universidad del Pa\'is Vasco, UPV- EHU - Bilbao, Spain}
\author{Dario Poletti}
\email{dario_poletti@sutd.edu.sg}
\affiliation{Science and Math Cluster, Singapore University of Technology and Design, 8 Somapah Road, 487372 Singapore}
\affiliation{Engineering Product Development, Singapore University of Technology and Design, 8 Somapah Road, 487372 Singapore}
\begin{abstract}
The high fidelity generation of strongly entangled states of many particles, such as cat states, is a particularly demanding challenge. One approach is to drive the system, within a certain final time, as adiabatically as possible, in order to avoid the generation of unwanted excitations. However, excitations can be generated also by the presence of dissipative effects such as dephasing. Here we compare the effectiveness of Local Adiabatic and the FAst QUasi ADiabatic protocols in achieving a high fidelity for a target superposition state both with and without dephasing. In particular we consider trapped ions set-ups in which each spin interacts with all the others with the uniform coupling strength or with a power-law coupling. In order to mitigate the effects of dephasing, we complement the adiabatic protocols with dynamical decoupling and we test its effectiveness.
The protocols we study could be readily implemented with state-of-the-art techniques.
\end{abstract}
\maketitle
\section{Introduction}
The possibility of generating many-body entangled states has important consequences in metrology \cite{LeibfriedWineland2005, GilchristMilburn2004, GiovannettiMaccone2004, StrobelOberthaler2014} and in quantum computation \cite{Nielsen2006}.
One approach to produce such states is to first prepare the system in a ground state easy to obtain with high fidelity, e.g. in the presence of a strong magnetic field, and then adiabatically transfer the state to the target ground state of a modified Hamiltonian, e.g. by ramping down the magnetic field.
However, in such approach, one encounters two main difficulties:
the first is the presence of small avoided crossings,
what makes it difficult to follow the ground state adiabatically without producing excitations in a finite time;
the second is the presence of sources of dissipation which may also excite the system.
To counter the first point, one could choose to evolve the system very slowly.
However, for practical applications, it would be ideal to be able to prepare target states in times as short  as possible.
Moreover, it is clear that the longer the preparation of a state takes,
the longer the dissipation will affect the system, thus driving it away from the target state.
It is therefore necessary to use a strategy that allows, simultaneously,
both to prepare a target state quickly,
reducing the possible excitations from the Hamiltonian driving,
and to protect the system from the effects of dissipation.

In our work we consider a system of spins coupled to each other via phonon-mediated interactions,
a very tunable model for a quantum simulator.
In fact, such a model can be realized both in cavity QED systems \cite{LerouxVuletic2010, RitschEsslinger2013}
and with trapped ions \cite{PorrasCirac2004, Kim2009, Islam2011, Islam2013}.
The fact that the interactions are mediated by the phonons significantly increases the tunability of the set-ups,
but it also introduces a source of dissipation,
which is the dephasing due to the phonons themselves.
The preparation of target cat states in a trapped ion set-up was recently studied in \cite{Safavi-Naini2018}.
In this work they studied, both theoretically and experimentally,
the evolution of ions in a Penning trap.
They prepared a system in a product state in the presence of a large magnetic field,
and then reduced the magnetic field to drive it to a cat state.
In particular, they showed that tuning the magnetic field following a so-called local adiabatic (LA) approach \cite{Richerme2013} could significantly improve the fidelity of the target state, compared to a linear or an exponential ramp.

Here we aim to continue this line of research by combining two techniques:
the use of a different method to design the time evolution of the magnetic field,
namely the fast quasi adiabatic (FAQUAD) protocol \cite{Martinez-Garaot2015},
and the use of dynamical decoupling \cite{ViolaLloyd1998, ViolaLloyd1999}
to tame the effects of dephasing.
We will also consider both the scenarios of trapped ions in a two-dimensional Penning trap \cite{Bohnet2016} which results in equal interaction between all spins,
and that of a linear Paul trap in which the interaction as a function of distance follows a power-law decay \cite{Kim2009, Islam2013}.

The manuscript is structured in the following manner:
In Sec. \ref{sec:protocols} we describe the methods of shortcuts to adiabaticity \cite{Torrontegui2013} to design the optimized adiabatic protocols that we will use.
In Sec. \ref{sec:model} we focus on a spin system with uniform all-to-all interactions,
and we study the effectiveness of LA and FAQUAD protocols both for unitary and dissipative evolution.
For the latter, we will also consider the effect of dynamical decoupling.
In Sec. \ref{sec:isingmodel} we focus on a spin system with power-law interactions,
and in Sec. \ref{sec:conclusions} we draw our conclusions.
\section{Local Adiabatic and FAQUAD protocols}
\label{sec:protocols}
As mentioned in the introduction, it is possible to reduce the amount of excitation in the prepared state by designing an appropriate protocol for the time-dependence of the Hamiltonian control parameters.
Here we give an introduction to the Local Adiabatic (LA) \cite{Richerme2013} and the FAst QUasi-ADiabatic (FAQUAD) \cite{Martinez-Garaot2015} protocols.
The main idea is to start from the adiabaticity condition which imposes that the change in a state should be much smaller when the energy gap between this states and another relevant state. This translates to
\begin{equation}
\hbar\left|\frac{\left\langle\psi_{a}(t) | \partial_{t} \psi_{b}(t)\right\rangle}{E_{a}(t)-E_{b}(t)}\right| \ll 1,
\end{equation}
where the $|\psi_i\rangle$ are two eigenstates, and the $E_i$ the corresponding eigenenergies. The relevant states to consider are, for our application, the ground state and the first excited state which is coupled by the changing Hamiltonian (which, due to symmetries, could for instance be the second excited state of the instantaneous Hamiltonian $H(t)$).
It is thus possible to distribute homogeneously in time the probability of transition between the two energy levels involved in the equation by imposing the following condition%
\begin{equation}
\hbar\left|\frac{\left|\psi_{a}(t)\right| \partial_{t} \psi_{b}(t) \rangle}{E_{a}(t)-E_{b}(t)}\right|=\hbar\left|\frac{\left\langle\psi_{a}(t)\left|\frac{\partial H}{\partial t}\right| \psi_{b}(t)\right\rangle}{\left[E_{a}(t)-E_{b}(t)\right]^{2}}\right|=c. \label{eq:adicond}
\end{equation}
Then, rewriting the time vector as a function of the control parameter, $t = t(B_{\mu})$ (we label the control parameter as $B_{\mu}$ because in the following the control parameter will be the magnetic field $B_{\mu}$, by ${\mu}$ we indicate a particular time-dependence/protocol to vary it), Eq.(\ref{eq:adicond}) gives the FAQUAD protocol
\begin{equation}
\label{FAQUAD}
\dot{B}_{F}=\mp \frac{c}{\hbar}\left|\frac{\left[E_{a}(B_{F})-E_{b}(B_{F})\right]^{2}}{\left\langle\psi_{a}(B_{F})\left|\frac{\partial H}{\partial B_{F}}\right| \psi_{b}(B_{F})\right\rangle}\right|.
\end{equation}
By simply integrating this equation, one could obtain the control parameter $B_F$ as a function of time \cite{Martinez-Garaot2015}. Given the size of the systems we will study, the protocols will be evaluated numerically.
The LA method, see \cite{Richerme2013}, stems from Eq.(\ref{FAQUAD}), with an additional assumption that simplifies the calculation of the protocol. The assumption is that $\left\langle\psi_{a}(B)\left|\frac{\partial H}{\partial B}\right| \psi_{b}(B)\right\rangle = 1$ for all times, which results in the equation for the parameter $B_L(t)$
\begin{equation}
\label{LA}
\dot{B}_{L}=\mp \frac{c}{\hbar}\left|\left[E_{a}(B_L)-E_{b}(B_L)\right]^{2}\right|
\end{equation}
Since FAQUAD uses the adiabaticity condition more accurately, the ensuing protocol distributes the loss of adiabaticity better along the time evolution.
In the following we will also consider a modification of the FAQUAD protocol. In Eq.(\ref{FAQUAD}) we only considered the ground state and the first relevant excited state. However, the time-dependent Hamiltonian could couple, in a non-negligible manner, the instantaneous ground state to a few excited levels.
We thus obtain
\begin{equation}
\label{FAQUAD_Multi}
\dot{B}_{K}=\frac{c}{\hbar} \left(\sum_{k=1}^K\left|\frac{\left\langle\psi_{g}(B_K)\left|\frac{\partial H}{\partial B_K}\right| \psi_{k}(B_K)\right\rangle}{\left[E_{g}(B_K)-E_{k}(B_K)\right]^{2}}\right| \right)^{-1},
\end{equation}
where the subscript $g$ indicates the ground state, while the $k=1\rightarrow 5$ enumerates the lowest excited states coupled to the instantaneous ground state.
In this paper we considered up to $K=5$ relevant transitions to design the protocols, and we refer to the FAQUAD protocols that consider $K$ relevant transitions as FAQUAD-K.

\section{Uniform all-to-all interactions}
\label{sec:model}
We consider $N$ ions in a trap with spins degrees of freedom coupled to a normal mode of the system via a spin-dependent optical dipole force.
In particular, we focus on the case in which the optical dipole force is tuned such that the center of mass mode is uniformly coupled to all the ions/spins. This setup, with the addition of an external magnetic field $B_{\mu}(t)$ can be described by the Dicke Hamiltonian \cite{Dicke1954, Garraway2011, WallRey2017}
\begin{equation}
\label{Dicke_Hamiltonian}
\hat{H}_{\text { Dicke }} = -\frac{\hbar g_{0}}{\sqrt{N}}\left(\hat{a}+\hat{a}^{\dagger}\right) \hat{S}_{z} + B_{\mu}(t) \hat{S}_{x} - \hbar\delta \hat{a}^{\dagger} \hat{a},
\end{equation}
where $a, a^\dagger$ are the bosonic annihilation and creation operators, $B_{\mu}(t)$ is the time-dependent transverse magnetic field from the protocol $\mu=F, L, K$ (see Eqs.(\ref{FAQUAD},\ref{LA},\ref{FAQUAD_Multi})),
$g_0$ is the coupling between the spins and the center of mass mode, and $\delta$ is the detuning between the optical dipole force and the center of mass mode. In Eq. \eqref{Dicke_Hamiltonian} we have used the collective operator $\hat{S}_{\alpha}=(1 / 2) \sum_{j} \hat{\sigma}_{j}^{\alpha}$, being $j$ the label for each spin, and $\hat{\sigma}_{j}^{\alpha}$ the Pauli matrices for $\alpha = x, y, z$.

%
%
At this point it is possible to rewrite the Hamiltonian \eqref{Dicke_Hamiltonian} as
\begin{equation}
\hat{H}(t)=-\hbar\delta \hat{b}^{\dagger} \hat{b}+\frac{J}{N} \hat{S}_{z}^{2}+B_{\mu}(t) \hat{S}_{x},
\end{equation}
where
$\hat{b}=\hat{a}-\left[g_{0} /(\sqrt{N} \delta)\right] \hat{S}_{z}$
and
$J = \hbar g_0^2/\delta$.
Here the term $\sim \hat{b}^{\dagger} \hat{b}$ describes the phonons in a displaced potential, with a spin dependent displacement.
In the limit of large detuning, $|\delta| \gg g_{0} / \sqrt{N}$, the bosonic mode can be adiabatically eliminated, leaving a purely spin system
\begin{equation}
\label{LM_Hamiltonian}
\hat{H}_{\mathrm{LM}}(t)=\frac{J}{N} \hat{S}_{z}^{2}+B_{\mu}(t) \hat{S}_{x},
\end{equation}
which, for $\delta<0$ is known as the ferromagnetic Lipkin model Hamiltonian.
The Lipkin and the Dicke Hamiltonians show a similar behavior:
for large magnetic field, the ground state is such that the spins are polarized in the $x$ direction, while for small magnetic field they are in a symmetric superposition of all spins pointing up plus all spins pointing down in the $z$ direction. However, in the latter case, the energy gap is small and it is thus difficult to prepare such state via adiabatic driving.

Using the Lipkin model to describe the physical setup has two important advantages that significantly simplify the study. First it eliminates the bosonic degree of freedom, and second it allows to further reduce the relevant Hilbert space as the spin states will belong to the Dicke manifold composed of $N+1$ states only, each a symmetric of superposition of the spin states with the same magnetization.

More precisely, for $N$ spins and $\bar n$ maximum bosonic occupation, the size of the relevant Hilbert space for the Dicke Hamiltonian in Eq. \eqref{Dicke_Hamiltonian} is $D_D = 2^N(\bar n+1)$, while for the Lipkin model Eq. \eqref{LM_Hamiltonian} it is $D_\mathrm{LM} = 2^N$. However, if the initial condition is in the symmetric sector in which all spins are prepared in the same state, we notice that the state will only evolve within this symmetry sector. Hence the dimensionality of the relevant Hilbert spaces can be reduced to $D_D^s = (N+1)(\bar n+1)$ and $D_L^s = N+1$.

The experimental realization of such a setup, however, comes with dissipative effects. Following \cite{Safavi-Naini2018}, we will consider that the open dynamics is mostly subject to dephasing, which can be described by a master equation of the Gorini-Kossakowski-Sudarshan-Lindblad form \cite{Lindblad1976, Gorini1976}
\begin{equation}
\label{master_equation}
\frac{d \hat{\rho}}{d t}=-\frac{\im}{\hbar}\left[\hat{H}, \hat{\rho}\right]+\frac{\Gamma}{2} \sum_{i=1}^{N}\left(\hat{\sigma}_{i}^{z} \hat{\rho} \hat{\sigma}_{i}^{z}-\hat{\rho}\right),
\end{equation}
such that the dephasing is produced by single site $\sigma^z$ operators.

\subsection{Closed system scenario}\label{sec:close}
We first consider an ideal case in which there is no dissipation, i.e. $\Gamma=0$.
We take the protocols designed using the FAQUAD and LA methods in Eqs. \eqref{FAQUAD} to \eqref{FAQUAD_Multi},
and calculate the corresponding time evolution solving the Schr\"odinger equation using a Runge-Kutta solver.
We start from the paramagnetic ground state for a large initial magnetic field ($B_0/(2\pi) = 7$ KHz),
and aim to reach the ferromagnetic superposition state at final time when $B_{\mu} = 0$.
As a representative case, we run a simulation for $N = 6$ spins in the Lipkin model \eqref{LM_Hamiltonian},
with parameters that can be reproduced in state-of-the-art labs \cite{Safavi-Naini2018}.
The dependence on time of the magnetic field for the FAQUAD and LA protocols are depicted in Fig. \ref{fig:fidelity_open_N6}(a) respectively by the blue solid line and the red dashed line.
The thick lines in Fig. \ref{fig:fidelity_open_N6}(b) depict the fidelities obtained after the evolution following the LA (dashed red line) and FAQUAD (solid blue line) approaches as a function of final time.
More precisely, we evolve the initial state with a protocol determined by the chosen final time, and for each of these final times we measure the final fidelity.
We define the fidelity as $\fid=\tr(\rho |\psi_{Target}\rangle\langle\psi_{Target}|)$ where $\rho$ is the actual state reached at final time $t_f$, and $\psi_{Target}$ is the superposition ground state of the Lipkin Hamiltonian at $B_{\mu}=0$ \cite{fn1}.

Unlike for the LA protocol, in which case the fidelity increases monotonously with the final time,
for FAQUAD-based protocols the fidelity shows an oscillatory behavior.
This implies that while FAQUAD allows to reach high fidelities at shorter times \cite{Martinez-Garaot2015}, it is also possible to find final times for which LA performs better. However, since the general objective is to obtain as good fidelities as possible in the shortest possible final time,
we can state that FAQUAD already brings a clear improvement with respect to LA.
For instance, if we fix a target fidelity of $\fid = 0.99$, we see that using FAQUAD protocol it is possible to reach such value at final times $t_f \sim 4.8$ ms, while using LA protocol only reaches that same fidelity only at $t_f = 12.4$ ms.
It is thus possible to reach the same level of fidelity in a time $2.6$ times shorter.
This result can be further improved, for instance, we have found that for the same number of spins, by varying the system parameters in the vicinity of the values used in Fig.\ref{fig:fidelity_open_N6}(b) FAQUAD can require a time which is even $4$ times shorter compared to LA.

\subsection{Performance of the unitary protocols in the open system scenario}
In experimental set-ups, dephasing can affect the fidelity in a detrimental way. Since the FAQUAD protocol [Eq. \eqref{FAQUAD}] can reach higher fidelities in a shorter time, it could potentially perform better than the LA protocol, Eq. \eqref{LA}.
However, one protocol could drive the state for longer periods to states more easily affected by dissipation, and thus result in worse performance. Here we analyze quantitatively the effect of dephasing on the resulting fidelity when using the FAQUAD or LA protocols described in Eqs.(\ref{FAQUAD},\ref{LA}).

Since the Lindblad dissipator acts independently on the local spins, unlike the unitary Dicke or Lipkin Hamiltonians
[Eqs. \eqref{Dicke_Hamiltonian} and \eqref{LM_Hamiltonian}], which are purely functions of the collective spins,
the evolution of the system cannot be solely described by the symmetric subspaces analyzed in Sec.\ref{sec:close}.
Due to this, in order to have an accurate description of the system evolution using the Dicke Hamiltonian \eqref{Dicke_Hamiltonian}, one would need a vector space of dimension $2^{2N}(\bar{n}+1)^2$, or of dimension $2^{2N}$ when using the Lipkin Hamiltonian \eqref{LM_Hamiltonian}.
In the following we will concentrate on the latter.
%
\begin{figure}[t]
\begin{center}
\includegraphics[width=\linewidth]{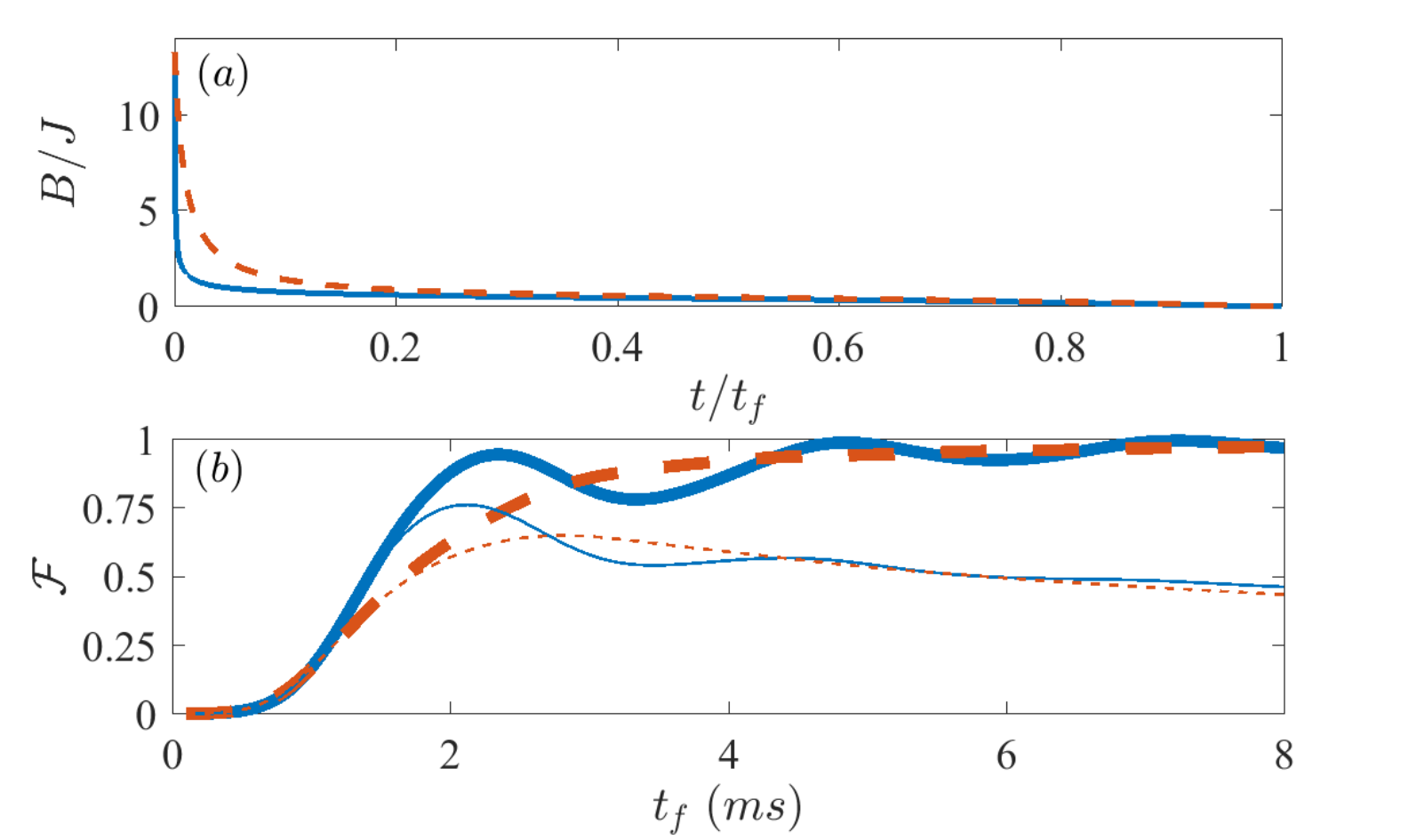}
\caption{
\label{fig:fidelity_open_N6}
Fidelity $\fid$ after evolving the master equation \eqref{master_equation} using the Lipkin Hamiltonian \eqref{LM_Hamiltonian}.
Solid blue lines are for FAQUAD and dashed red lines for LA. Thick lines show the fidelities for the unitary evolution, i.e. $\Gamma ~=~ 0$ s$^{-1}$, whereas the thin lines are for the open system with dephasing $\Gamma ~=~ 120$ s$^{-1}$.
The inset shows the protocols derived from Eqs. (\ref{FAQUAD}, \ref{LA}).
The rest of the parameters are, $N = 6$, $B_0 /(2\pi) = 7$ KHz, $J = 0.55N$ KHz, and $\delta/(2\pi) = -4$ KHz.
}
\end{center}
\end{figure}
%

We apply again the theory in Sec. \ref{sec:protocols} for the Lipkin model \eqref{LM_Hamiltonian},
and use it to solve the open dynamics described by the master equation \eqref{master_equation}.
While the protocols are computed only taking into account the unitary part of the master equation, if they are effective over a short evolution they should also be less affected by dissipation. In particular, since dephasing is a cumulative effect, we expect that at short times the protocols will perform similarly to the closed system scenario. For longer times instead, we expect a decay in the fidelity caused by dephasing.
The thin lines in Fig. \ref{fig:fidelity_open_N6}(b) show precisely this, where the results with protocols from FAQUAD are depicted by a thin blue solid line, while from LA by the thin red dashed line.
At short final times $t_f$, the evolution of fidelity versus $t_f$ is identical for the dynamics in a closed (thick lines) or open system scenario (thin lines), but at around $2$ms the respective curves diverge and the peak of fidelity for open systems occurs at much lower values of $t_f$.
After this maximum, the fidelity for the open system case decays almost monotonously vs the final time, whereas the one for the closed system approaches unit fidelity. For the parameters simulated in Fig. \ref{fig:fidelity_open_N6}(b), the maximum fidelity with the FAQUAD protocol is $\fid_{F} = 0.7628$, whereas the maximum fidelity for the LA protocol is $\fid_{LA} = 0.6514$, which implies that the use of the FAQUAD protocol results in a $11\%$ increase in fidelity.

%
\begin{figure}[t]
\begin{center}
\includegraphics[width=\linewidth]{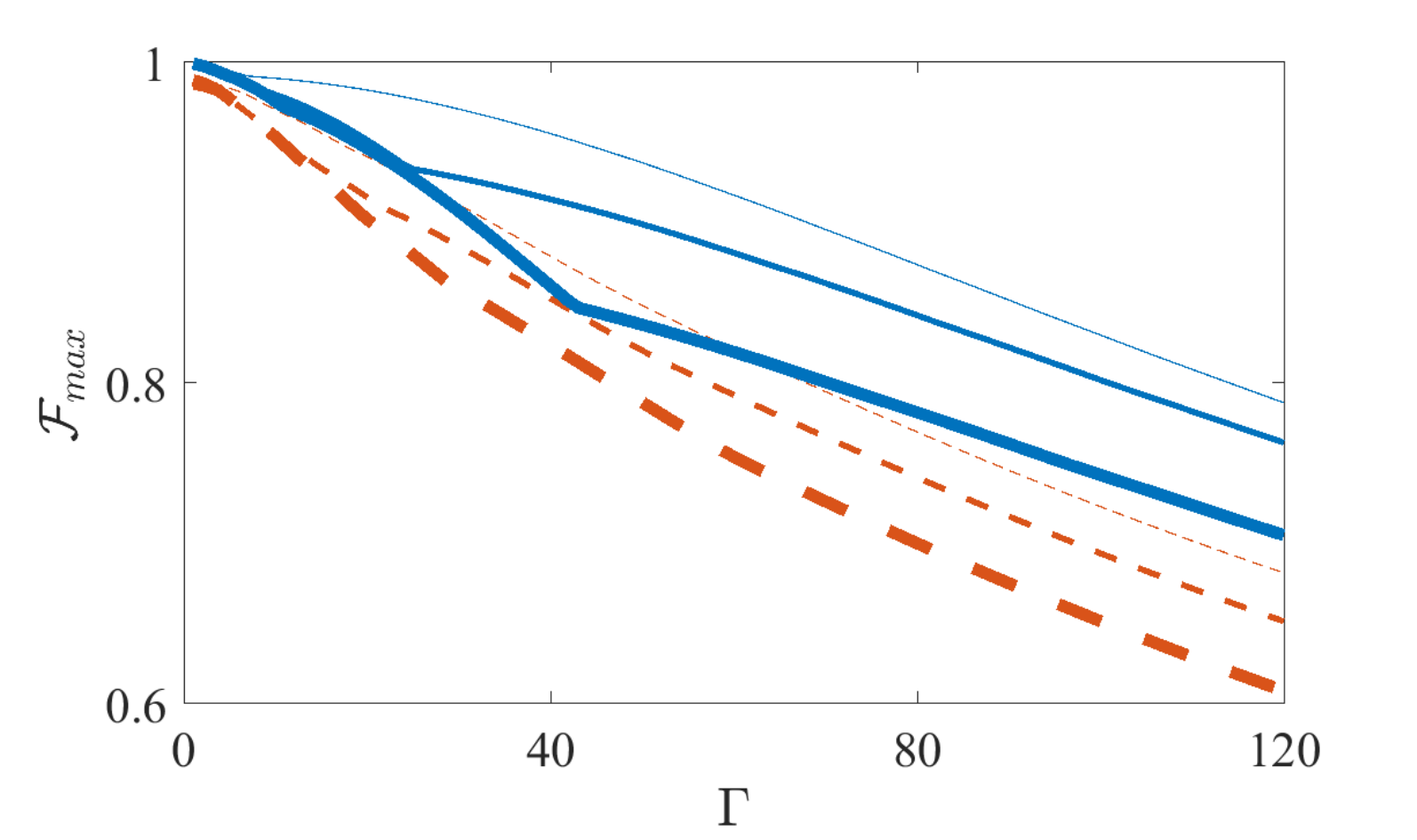}
\caption{
\label{fig:LipkinvsGamma}
Maximum fidelity $\fid_{max}$ vs the value of the dephasing $\Gamma$ after evolving the master equation \eqref{master_equation} using the Lipkin Hamiltonian \eqref{LM_Hamiltonian}.
Solid blue lines are for FAQUAD and dashed red lines for LA,
and the increasing thickness of the lines mean increasing system size, namely $N = 4$, $N = 6$ and $N = 8$.
The remaining parameters are, $B_0 /(2\pi) = 7$ KHz, $J = 0.55N$ KHz, and $\delta/(2\pi) = -4$ KHz.
}
\end{center}
\end{figure}
%
In Fig. \ref{fig:LipkinvsGamma} we show how the value of the maximum fidelity attainable $\fid_{max}$ depends on the magnitude of dephasing for different system sizes and on the protocol used, in particular FAQUAD [Eq. \eqref{FAQUAD}] or LA [Eq. \eqref{LA}].
In Fig. \ref{fig:LipkinvsGamma} we notice that for a given system size, protocols from FAQUAD (blue solid lines), perform better than from LA (red dashed lines). However, we also observe that as the system size increases ($N=4,\;6$ and $8$, from thinner to thicker lines) the maximum fidelity decreases.
\subsection{Dynamical decoupling of the dephasing}
%
\begin{figure}[t]
\begin{center}
\includegraphics[width=\linewidth]{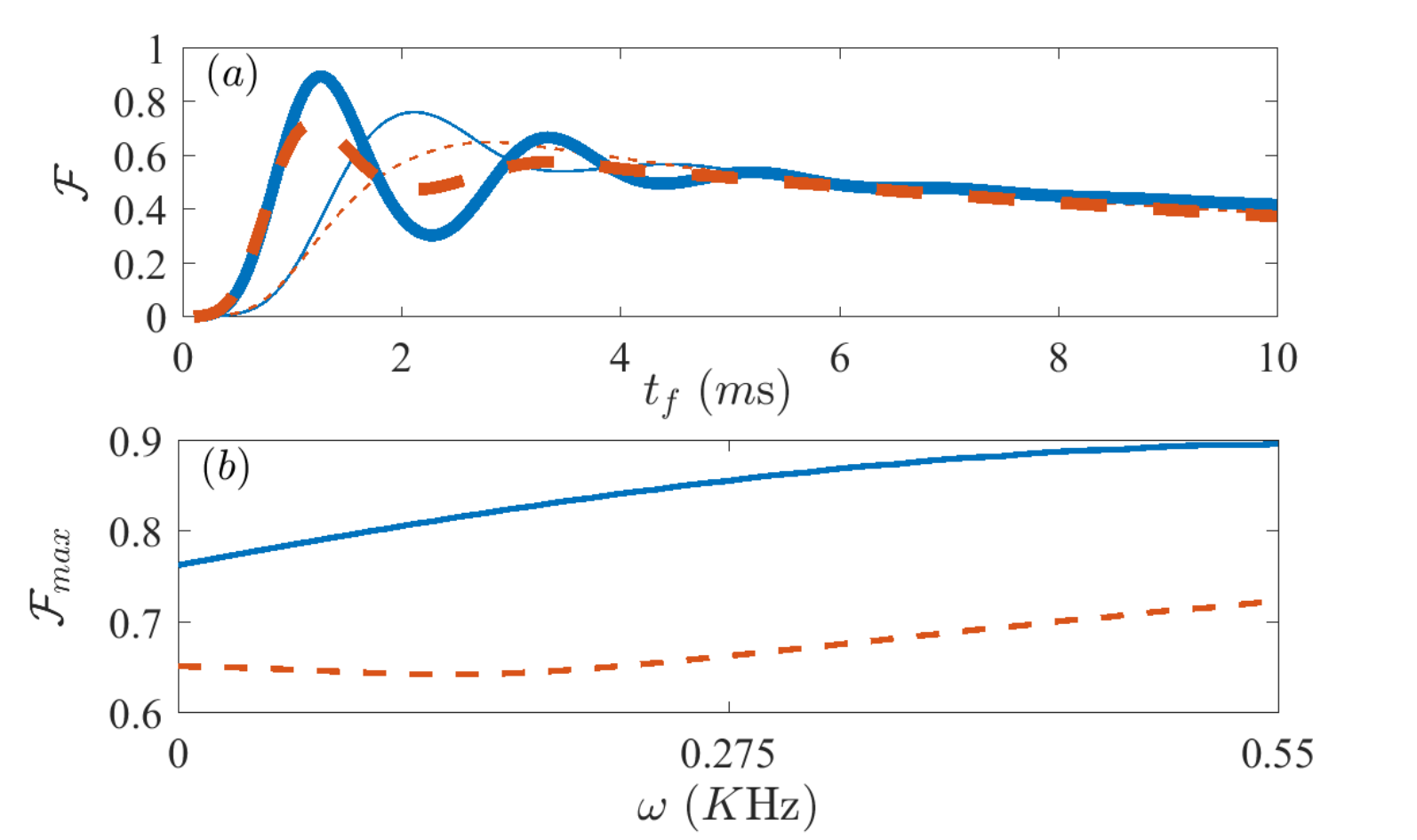}
\caption{
\label{fig:fidelity_DD}
Fidelity $\fid$ after evolving the master equation \eqref{master_equation} using the dynamically decoupled Lipkin Hamiltonian \eqref{LM_Hamiltonian_DD}.
(a) shows the maximum value of the fidelity at the first peak for each value of $\omega$ between 0 KHz and 0.55 KHz,
and (b) displays the fidelity vs final time $t_f$ for the value of $\omega$ with best absolute fidelity (thick lines) and for the case without dynamical decoupling (thin lines).
Both panels compare results for the FAQUAD (solid blue) and LA (dashed red) protocols.
Parameters are, $N = 6$, $B_0 /(2\pi) = 7$ KHz, $J = 0.55N$ KHz, and $\delta/(2\pi) = -4$ KHz, $\Gamma ~=~ 120$ s$^{-1}$.
}
\end{center}
\end{figure}
%

In \cite{ViolaLloyd1998, ViolaLloyd1999} it was shown that the use of additional terms to the Hamiltonian could result in filtering out unwanted effects of the system-bath interactions, what is known as ``dynamical decoupling''.
In this case, because in our effective Hamiltonian \eqref{LM_Hamiltonian} we have a term proportional to $S_z^2$,
we will add a dynamical decoupling term, proportional to $S_y^2$.
For the Lipkin Hamiltonian with dephasing, and inspired by \cite{Sun2016}, we notice that the addition of a term proportional to $S_y^2$ can be effective for dynamical decoupling.

To effectively obtain such $S_y^2$ term, we propose to introduce an independent spin-boson field using a new pair of lasers. The Dicke Hamiltonian with the new term ($\hat{c}$) will take the form
\begin{align}
\hat{H}'_{\text{Dicke}}/\hbar  = &-\frac{g_{0}}{\sqrt{N}}\left(\hat{a}+\hat{a}^{\dagger}\right) \hat{S}_{z}+B_{\mu}(t) \hat{S}_{x}-\delta \hat{a}^{\dagger} \hat{a} \nonumber\\
&-\frac{g_{0}}{\sqrt{N}}\left(\hat{c}+\hat{c}^{\dagger}\right) \hat{S}_{y}-\delta' \hat{c}^{\dagger} \hat{c}.
\end{align}
If we now rewrite the bosonic mode as in Sec. \ref{sec:model}, we get
\begin{align}
\hat{H}(t) = &-\delta \hat{b}^{\dagger} \hat{b}+\frac{J}{N} \hat{S}_{z}^{2}+B(t) \hat{S}_{x}\nonumber\\
& -\delta \hat{d}^{\dagger} \hat{d}+\frac{J'}{N} \hat{S}_{y}^{2},
\end{align}
where, as in Sec. \ref{sec:model}, $\hat{b}=\hat{a}-\left[g_{0} /(\sqrt{N} \delta)\right] \hat{S}_{z}$ and $J = \hbar g_0^2/\delta$, and for the additional introduced field $\hat{d}=\hat{c}-\left[g_{0} /(\sqrt{N} \delta')\right] \hat{S}_{y}$ and $J' = \hbar g_0^2/\delta' \equiv N\omega  \sin\left(\frac{\pi t}{t_f} \right)$.
We introduced the final equivalence so that the dynamical decoupling term will have an optimizing constant $\omega$, and we chose a sinusoidal time-dependence so that the dynamical decoupling term will be zero at initial and final times. The large detuning limit for the new field will require $\omega  \sin\left(\frac{\pi t}{t_f} \right) \ll \frac{g_0}{\sqrt{N}}$, so for a small $\omega$ we can always rewrite the Hamiltonian in the Lipkin model form
\begin{equation}
\label{LM_Hamiltonian_DD}
\hat{H}'_{\mathrm{LM}}(t)=\frac{J}{N} \hat{S}_{z}^{2}+\omega  \sin\left(\frac{\pi t}{t_f} \right)\hat{S}_{y}^2+B(t) \hat{S}_{x}.
\end{equation}

We use this Hamiltonian in Fig. \ref{fig:fidelity_DD},
mapping the optimizing parameter in the range $\omega\in [0, 0.55]$ KHz.
In Fig. \ref{fig:fidelity_DD}(a) we show the fidelity vs final time $t_f$, both without dynamical decoupling ($\omega=0$ KHz, thin lines) and with dynamical decoupling ($\omega= 0.55$ KHz), both for FAQUAD (solid blue lines) and for LA (red dashed lines).
In Fig. \ref{fig:fidelity_DD}(b) we plot the maximum value of the fidelity at the highest peak vs the optimizing parameter $\omega$.
Again the blue solid line reflects the results for FAQUAD, while the red dashed line those for LA.
Here we observe that the maximum fidelity we can obtain for $\omega = 0.55$ KHz is  $\fid_{\mathrm{FAQ}} = 0.8968$ for the FAQUAD protocol, a $13.5\%$ improvement, and $\fid_{\mathrm{LA}} = 0.7234$ for LA, corresponding to a $7\%$ improvement. We note that we chose not to explore beyond $\omega = 0.55$ KHz because otherwise the Hamiltonian that we use may not  satisfying the condition to adiabatically eliminate the bosonic mode [see discussion before Eq. \eqref{LM_Hamiltonian_DD}].

%
\begin{figure}[t]
\begin{center}
\includegraphics[width=\linewidth]{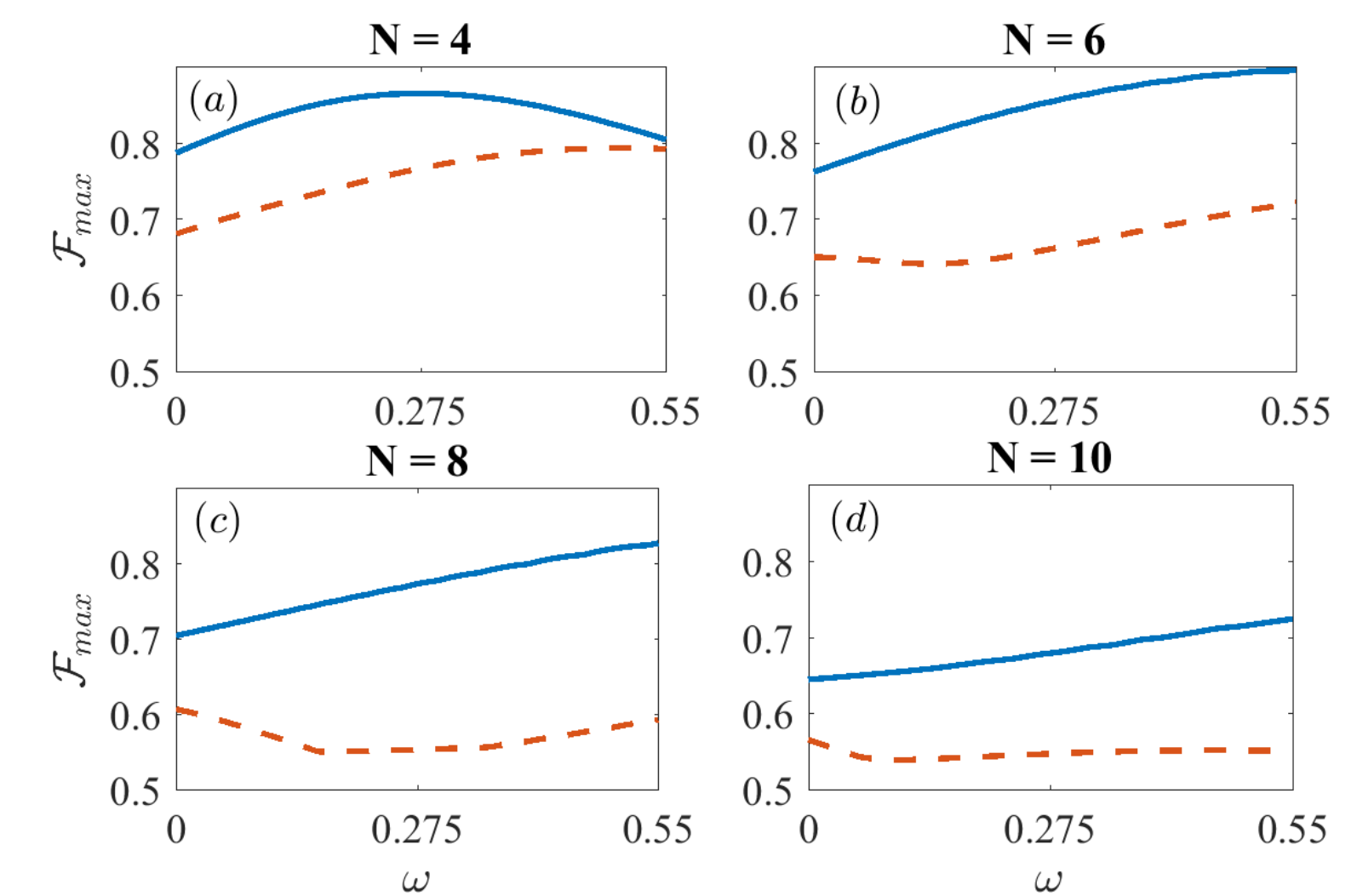}
\caption{
\label{fig:size}
Maximum value of the fidelity after evolving the master equation \eqref{master_equation} using the dynamically decoupled Lipkin Hamiltonian \eqref{LM_Hamiltonian_DD} vs parameter $\omega$.
We compare the fidelities for the FAQUAD (solid blue) and LA (dashed red) protocols for different chain sizes, ranging from $N = 4$ to $N = 10$.
Parameters are, $B_0 /(2\pi) = 7$ KHz, $J = 0.55N$ KHz, and $\delta/(2\pi) = -4$ KHz, $\Gamma ~=~ 120$ s$^{-1}$.
}
\end{center}
\end{figure}
%
In Fig. \ref{fig:size}, we study the effect of the chain size. Similarly to Fig. \ref{fig:fidelity_DD} we plot the maximum fidelity $\fid_{max}$ vs the dynamical decoupling magnitude $\omega$ for chains of size ranging between $N = 4$ and $N = 10$.
As the size grows, the maximum fidelity with no dynamical decoupling decreases with the system size.
However, for the FAQUAD protocols studied we observe that the fidelity can (sometimes significantly) be increased thanks to dynamical decoupling.
\section{Power-law interactions}
\label{sec:isingmodel}

Until now we have considered the scenario in which the interaction between the spins is uniform.
In trapped ions setups the interaction can also be of power-law form with a tunable exponent.
For instance, in \cite{Islam2011, Islam2013} the authors were able to realize the long range Ising Hamiltonian
\begin{equation}
\label{Ising_Hamiltonian}
\hat{H}_{\mathrm{Ising}}=\sum_{i<j} J_{i, j} \hat{\sigma}^{x}_{i} \hat{\sigma}^{x}_{j}+B_{\mu}(t) \sum_{i} \hat{\sigma}^{y}_{i},
\end{equation}
where $J_{i, j} = \frac{J_{max}}{\left| i-j\right|^\alpha}$.
As mentioned, the exponent of the long range interaction $\alpha$ can be tuned within a certain range.
Here we will consider the two values $\alpha = 1.2$ and $\alpha = 0$,
the latter for comparison purposes with the previous results using the Lipkin Hamiltonian \eqref{LM_Hamiltonian}.
For a clearer comparison with previous results, we will choose similar values of the parameters, so that the transverse magnetic field $B_{\mu}$ will decay from an initial value $B_{\mu}(0)/(2\pi) = 7$ KHz to a final value $B_{\mu}(t_f) = 0$ KHz, and we will also choose a similar interaction $J_{max} = -0.55$ KHz.
Similar to the Spin-Boson model, in the presence of a dominant transverse field,
the ground state is initialized with all spins aligned in the $y$ direction.
As the magnetic field is decreased to $0$, the ground state becomes a degenerate ferromagnetic state (we consider $J$ to be negative) in the $x$ direction.

To account for the external noise, in this system we will consider a local dephasing $\hat{\sigma}^x_i$ which is in the same direction of the spin-spin interaction, with a master equation
\begin{equation}
\label{master_equation_Ising}
\frac{d \hat{\rho}}{d t}=-\frac{\im}{\hbar}[\hat{H}_\mathrm{Ising}, \hat{\rho}]+\frac{\Gamma}{2} \sum_{i=1}^{N}\left(\hat{\sigma}_{i}^{x} \hat{\rho} \hat{\sigma}_{i}^{x}-\hat{\rho}\right) .
\end{equation}
We will study the performance of the LA and FAQUAD protocols designed for the unitary Hamiltonian \eqref{Ising_Hamiltonian}.
As we did before for the Lipkin model, we will also add a dynamically decoupling term.
For this case we will add an oscillating term proportional $\hat{\sigma}^z_i\hat{\sigma}^z_{j}$ such that
\begin{align}
\label{Ising_Hamiltonian_DD}
\hat{H}_{\text{Ising}}' &= \sum_{i<j} J_{i, j} \hat{\sigma}^{x}_{i} \hat{\sigma}^{x}_{j}+B_{\mu}(t) \sum_{i} \hat{\sigma}^{y}_{i}\nonumber\\
&+\omega \sin \left(\frac{\pi t}{t_{f}}\right)\sum_{i<j} \frac{1}{\left| i-j\right|^{\widetilde{\alpha}} } \hat{\sigma}^{z}_{i} \hat{\sigma}^{z}_{j},
\end{align}
where we consider in principle a different decay rate $\widetilde{\alpha}$ as for the unitary Hamiltonian.
%
\begin{figure}[t]
\begin{center}
\includegraphics[width=\linewidth]{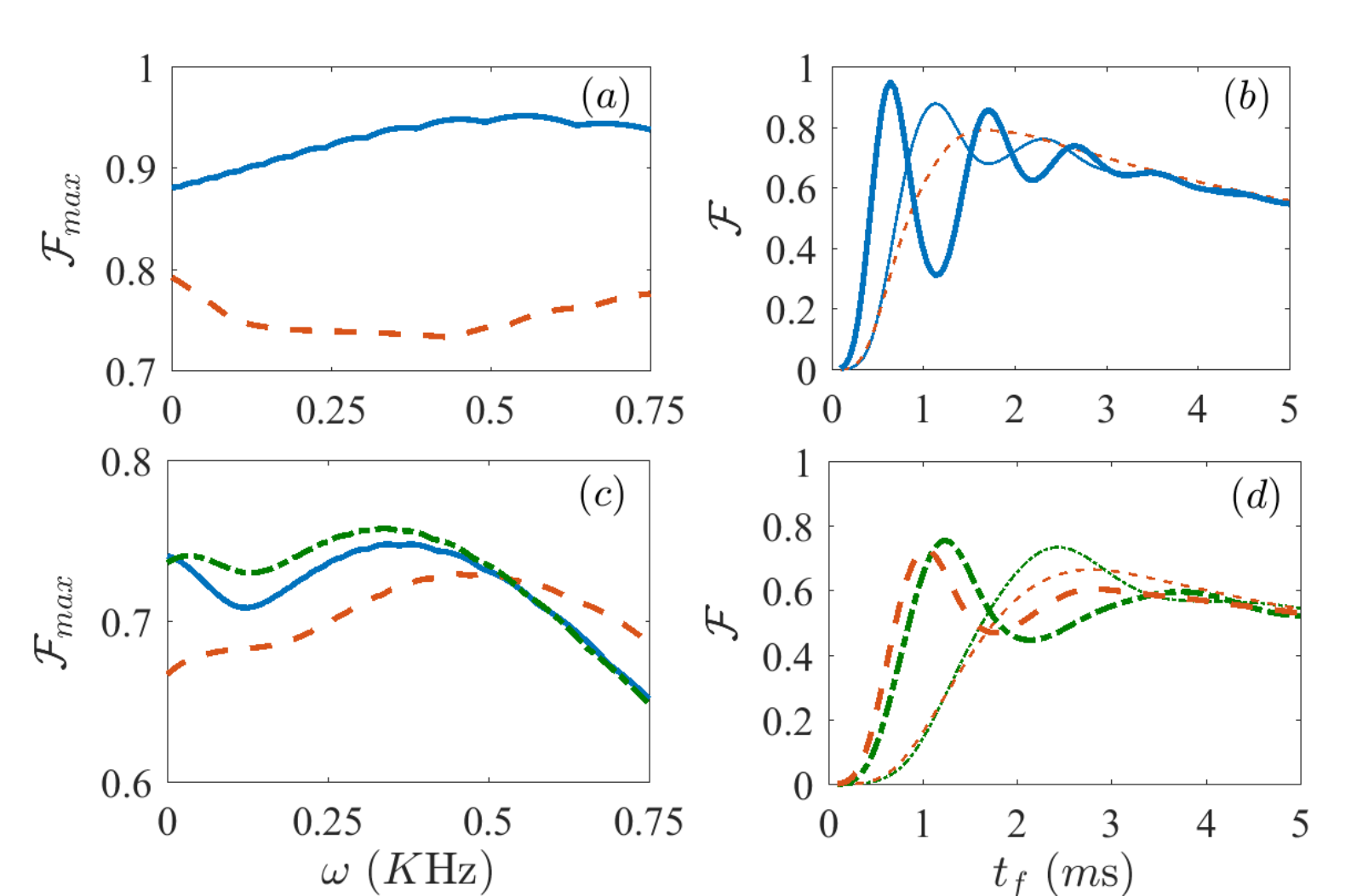}
\caption{
\label{fig:fidelity_ising_DD}
Fidelity $\fid$ after evolving the master equation \eqref{master_equation_Ising} using the dynamically decoupled Ising Hamiltonian \eqref{Ising_Hamiltonian_DD} for $\alpha = 0$ (upper panels) and $\alpha = 1.2$ (lower panels). (a) and (c) show the maximum value of the fidelity at the first peak for each value of $\omega$ between 0 KHz and 0.75 KHz, while (b) and (d) display the fidelity vs final time $t_f$ for the value of $\omega$ with best absolute fidelity (thick lines) and for the case without dynamical decoupling (thin lines). In panels (a-c) the solid blue lines are results for FAQUAD and dashed red lines for LA. In panel (c) we additionally have dash-dotted yellow lines for FAQUAD-4, the FAQUAD protocol that considers up to 4 relevant energy level transitions. In panel (d), solid blue lines are for FAQUAD4, and dashed red lines for LA. Parameters are, $N = 6$, $B_0 /(2\pi) = 7$ KHz, $J_{max} = 0.55$ KHz, $\widetilde{\alpha} = 0$, and $\Gamma ~=~ 120$ s$^{-1}$.
}
\end{center}
\end{figure}
%

%
\begin{figure}[t]
\begin{center}
\includegraphics[width=\linewidth]{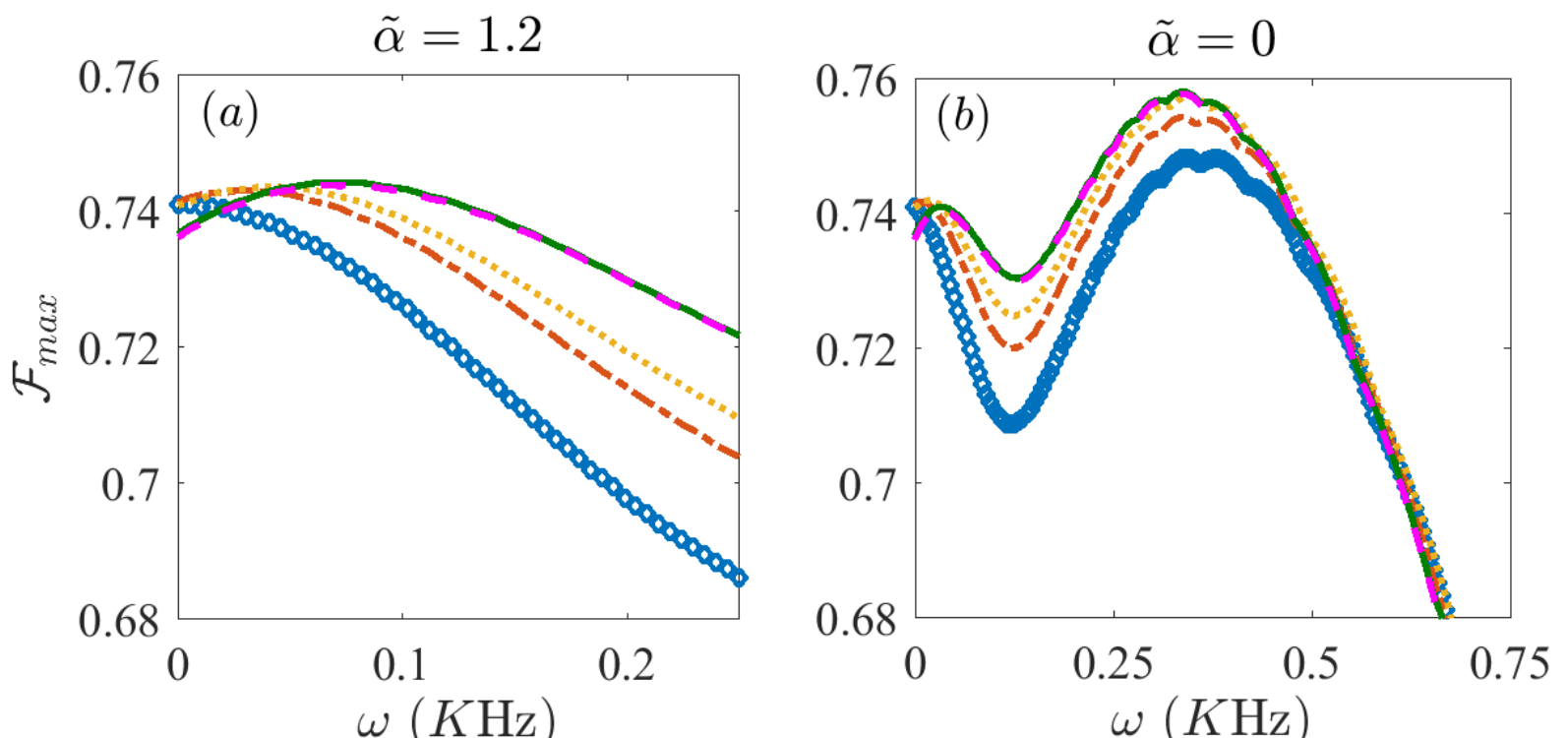}
\caption{
\label{fig:Fig6}
Maximum fidelity $\fid_{max}$ vs $\omega$ for the FAQUAD protocols considering a different number of meaningful transitions for $\widetilde{\alpha} = 1.2$ in panel (a) and $\widetilde{\alpha} = 0$ in panel (b). Blue diamonds are for FAQUAD-1, red dash-dotted line for FAQUAD-2, yellow dotted line for FAQUAD-3, solid green line for FAQUAD-4 and magenta dashed line for FAQUAD-5. Parameters are, $N = 6$, $B_0 /(2\pi) = 7$ KHz, $J_{max} = 0.55$ KHz, and $\Gamma ~=~ 120$ s$^{-1}$.
}
\end{center}
\end{figure}
%

In Fig.~\ref{fig:fidelity_ising_DD} we depict the fidelities after the evolutions using LA and FAQUAD protocols obtained for the unitary Hamiltonian \eqref{Ising_Hamiltonian}, with and without a dynamically decoupling term. In panels (a,c) we plot the maximum fidelity vs the magnitude of dynamical decoupling $\omega \in [0, 0.75]$ KHz for FAQUAD (blue solid line), LA (red dashed line) and FAQUAD-4 (yellow dot-dashed line).
In panels (b,d) we show only the results corresponding to no dynamical decoupling (thin lines) and with $\omega$ corresponding to the highest fidelity (note that for panel (b) the maximum fidelity for LA is obtained already for $\omega=0$).
Panels (a,b) are for $\alpha=0$, while panels $(c,d)$ for $\alpha=1.2$, i.e. with a power-law potential.
For $\alpha = 0$, the best fidelities obtained before applying dynamical decoupling with a dephasing $\Gamma ~=~ 120$ s$^{-1}$ are
$\fid_F = 0.8813$ and $\fid_L = 0.7234$, about $16\%$ better for the FAQUAD protocol.
Once we introduce the dynamical decoupling term, the FAQUAD protocol can reach a maximum fidelity up to a $7\%$ higher
$\fid_{F} = 0.9523$ when $\omega = 553.7$ Hz, whereas the result for LA does not improve for the parameters explored.

Until this point, we have considered uniform all-to-all interactions, and for all these cases, the use of FAQUAD-K, see Eq.(\ref{FAQUAD_Multi}), would not result in sizeable improvements in the fidelities. However, with a decaying long range interaction with $\alpha = 1.2$, we have tested for up to 5 relevant levels and the best fidelities are obtained for the FAQUAD protocol that considers up to $4$ meaningful level transitions, i.e. FAQUAD-4 (more details given in the discussion of Fig. \ref{fig:Fig6}).
Without dynamical decoupling, the highest fidelities are $\fid_{4} = 0.7367$ and $\fid_L = 0.6674$, almost a $7\%$ better for the FAQUAD-4 protocol.
Introducing dynamical decoupling, the best results are obtained for $\omega = 337.25$ Hz and $\omega = 448$ Hz respectively, reaching maximum fidelities of $\fid_{4} = 0.7582$, $2\%$ improvement, and $\fid_L = 0.7298$, $6\%$ improvement.

The analysis of the dynamical decoupling for different types of protocols FAQUAD-K, for $K=1$ to $5$ is done in Fig. \ref{fig:Fig6}.
In Fig.~\ref{fig:Fig6}(a) we show the comparison of the maximum fidelity of the different FAQUAD-K protocols vs the parameter $\omega$ when the interaction in the dynamically decoupling term $\tilde{\alpha}$ is the same as the interaction in the unitary Hamiltonian, i.e. $\tilde{\alpha} = \alpha= 1.2$.
In Fig.~\ref{fig:Fig6}(b) we consider the case in which the spatial dependence of the interaction is different for the dynamical decoupling compared to the interaction term. In fact we compare the performance of the different FAQUAD-K protocols when $\tilde{\alpha} = 0$, an analysis closely connected to Fig. \ref{fig:fidelity_ising_DD}(c).
Interestingly, for both $\tilde{\alpha}=1.2$ and $\tilde{\alpha}=0$, FAQUAD-K protocols with larger K reach a smaller maximum fidelity ($\fid_{max}$) in the absence of dynamical decoupling ($\omega=0$), but they perform much better with the dynamical decoupling term in Eq.(\ref{Ising_Hamiltonian_DD}). By comparing panels (a) and (b) in Fig. \ref{fig:Fig6} we also can clearly observe that in this case a non decaying interaction for the dynamical decoupling performs, for this set-up and parameters, better than implementing the same space-dependence of the interaction as in the unitary Hamiltonian.

\section{Conclusions}
\label{sec:conclusions}

We have studied the effectiveness of different protocols in producing cat states with good fidelity. We have considered setups that can be realized experimentally with trapped ions, both with uniform and power-law interactions. We have shown that FAQUAD protocols perfom better than LA protocols in providing final states with high fidelity in shorter times. Moreover, we have shown how important this is when considering the effect of dephasing too.
In fact, since FAQUAD protocols result in higher fidelities at short final times, the system has been under the influence of dephasing for a shorter period.
This improvement is specially notable in the uniform interaction case. For instance, for the study case in the Lipkin model we observed an improvement of an $11\%$ in fidelity. In the power law interaction case, we observed an improvement of $7\%$ with respect to the LA protocol.

We were also able to further improve the fidelity of the target states by introducing an additional field perpendicular to the coupling to the bath to dynamically decouple the system from the environment.
Notably, the presence of this additional term improves the fidelity for both FAQUAD and LA based protocols. In the cases studied here, we obtain an increase in fidelty up to $13\%$ fidelity.
Additionally, these larger maximum fidelities are reached at even shorter final times.

We have also considered models with spin interaction decaying as a power law, instead of a uniform all-to-all coupling. In this cases, we have observed that higher maximum fidelities can be reached with FAQUAD-K protocols, which is an extension of the FAQUAD protocol that takes into account the first $K$ relevant excited states. We have observed that protocols with higher $K$ can lead to an important improvement of the performance, especially in presence of dynamical decoupling.
Interestingly, the dependence in space of the interaction for the dynamical decoupling term could be different from that of the spin-interactions in the Hamiltonian, and we have observed that a more uniform interaction in the dynamical decoupling term could help increase the fidelity.

In future works we could study larger system sizes.
While the simulation of the dissipative dynamics is particularly demanding for a large number of spins, the computation of the protocol for the magnetic field only depends on the Hamiltonian (e.g. unitary evolution), and hence one can compute it for larger system sizes so that it could be tested in experiments.

{\bf Acknowldgements:} D.P. and M.P. are grateful to J.J. Bollinger, R.J. Lewis-Swan, A.M. Rey and A. Safavi-Naini for fruitful discussions. D.P. and M.P. acknowledge support from the Singapore Ministry of Education, Singapore Academic Research Fund Tier-II (project MOE2018-T2-2-142). M.A.S. acknowledges support by the Basque Government predoctoral program (Grant No. PRE-2018-2-0177).

\bibliography{ref}

\begin{thebibliography}{25}
\bibitem[{fn1()}]{fn1}
\bibinfo{note}{We run simulations for up to FAQUAD-5, but the fidelities for
  protocols taking more meaningful transitions into account did not, in this
  case, significantly improve the results for the simple FAQUAD, therefore, for
  clarity of exposition, we only display the results for FAQUAD and LA.}
\end{thebibliography}
\end{document}